\documentclass[sigconf]{acmart}

\copyrightyear{2026}
\acmYear{2026}
\setcopyright{cc}
\setcctype{by}

\acmConference[WWW '26] {Proceedings of the ACM Web Conference 2026}{April 13--17, 2026}{Dubai, United Arab Emirates.}
\acmBooktitle{Proceedings of the ACM Web Conference 2026 (WWW '26), April 13--17, 2026, Dubai, United Arab Emirates}
\acmISBN{979-8-4007-2307-0/2026/04}
\acmDOI{10.1145/3774904.3792575}
\usepackage{algorithm}
\usepackage{algorithmic}

\usepackage{makecell}
\usepackage{multirow}
\usepackage{array}
\usepackage{amsmath}
\usepackage{amsthm}
\usepackage{booktabs}
\usepackage{graphicx}
\usepackage{bm}
\usepackage{tabularx}
\usepackage{subcaption}
\usepackage{enumitem}
\usepackage[table]{xcolor}
\usepackage{xcolor}
\usepackage{xspace}

\newcommand{\model}{FCUCR\xspace}

\newcommand{\eg}{\emph{e.g.,}\xspace}
\newcommand{\ie}{\emph{i.e.,}\xspace}

\begin{document}

%%
%% The "title" command has an optional parameter,
%% allowing the author to define a "short title" to be used in page headers.

\title{Learning Evolving Preferences: A Federated Continual Framework for User-Centric Recommendation}

%%
%% The "author" command and its associated commands are used to define
%% the authors and their affiliations.
%% Of note is the shared affiliation of the first two authors, and the
%% "authornote" and "authornotemark" commands
%% used to denote shared contribution to the research.
\author{Chunxu Zhang}
\authornote{Both authors contributed equally to this research.}

\affiliation{%
  \institution{College of Computer Science and Technology, Jilin University}
  \institution{Key Laboratory of Symbolic Computation and Knowledge Engineering of Ministry of Education, Jilin University}
  \city{Changchun}
  \country{China} \\
  \institution{PolyU Academy for Artificial Intelligence, Hong Kong Polytechnic University}
  \city{Hong Kong}
  \country{China}
  }
\email{zhangchunxu@jlu.edu.cn}
  
\author{Zhiheng Xue}
\authornotemark[1]
\affiliation{%
  \institution{College of Computer Science and Technology, Jilin University}
  \institution{Key Laboratory of Symbolic Computation and Knowledge Engineering of Ministry of Education, Jilin University}
  \city{Changchun}
  \country{China}
  }
\email{xuezh24@mails.jlu.edu.cn}
  
\author{Guodong Long}
\affiliation{%
  \institution{Australian Artificial Intelligence Institute, FEIT, University of Technology Sydney}
  \city{Sydney}
  \country{Australia}
  }
\email{Guodong.Long@uts.edu.au}

\author{Weipeng Zhang}
\affiliation{%
  \institution{College of Computer Science and Technology, Jilin University}
  \institution{Key Laboratory of Symbolic Computation and Knowledge Engineering of Ministry of Education, Jilin University}
  \city{Changchun}
  \country{China}
  }
\email{zhangwp24@mails.jlu.edu.cn}

\author{Bo Yang}
\authornote{Corresponding author.}
\affiliation{%
  \institution{College of Computer Science and Technology, Jilin University}
  \institution{Key Laboratory of Symbolic Computation and Knowledge Engineering of Ministry of Education, Jilin University}
  \city{Changchun}
  \country{China}
  }
\email{ybo@jlu.edu.cn}

%%
%% By default, the full list of authors will be used in the page
%% headers. Often, this list is too long, and will overlap
%% other information printed in the page headers. This command allows
%% the author to define a more concise list
%% of authors' names for this purpose.
\renewcommand{\shortauthors}{Chunxu Zhang, Zhiheng Xue, Guodong Long, Weipeng Zhang, \& Bo Yang}

%%
%% The abstract is a short summary of the work to be presented in the
%% article.
\begin{abstract}
 User-centric recommendation has become essential for delivering personalized services, as it enables systems to adapt to users’ evolving behaviors while respecting their long-term preferences and privacy constraints. Although federated learning offers a promising alternative to centralized training, existing approaches largely overlook user behavior dynamics, leading to temporal forgetting and weakened collaborative personalization. In this work, we propose \model, a federated continual recommendation framework designed to support long-term personalization in a privacy-preserving manner. To address temporal forgetting, we introduce a time-aware self-distillation strategy that implicitly retains historical preferences during local model updates. To tackle collaborative personalization under heterogeneous user data, we design an inter-user prototype transfer mechanism that enriches each client’s representation using knowledge from similar users while preserving individual decision logic. Extensive experiments on four public benchmarks demonstrate the superior effectiveness of our approach, along with strong compatibility and practical applicability. Code is available at \url{https://github.com/Poizoner/code4FCUCR_www2026}.

 % \noindent\textbf{Relevance Statement:} This paper presents a federated continual recommendation framework that preserves user privacy while enabling long-term personalization. By tackling the challenges of temporal forgetting and collaborative personalization, this work is well aligned with the \textbf{user modeling, personalization, and recommendation} track.
\end{abstract}

%%
%% The code below is generated by the tool at http://dl.acm.org/ccs.cfm.
%% Please copy and paste the code instead of the example below.
%%



% \received{20 February 2007}
% \received[revised]{12 March 2009}
% \received[accepted]{5 June 2009}

%%
%% This command processes the author and affiliation and title
%% information and builds the first part of the formatted document.
\maketitle

\section{Introduction}
Modern recommender systems have become deeply embedded in our daily lives, shaping how users engage and make consumption decisions~\cite{hasan2024based,ge2024survey,li2024recent,liu2024multimodal}. However, existing cloud-based methods rely on periodic retraining using globally collected data. This strategy often fails to capture shifting user preferences in time. As a result, it leads to outdated models, forgetting past behaviors, and insufficient personalization. These limitations call for \textbf{a user-centric recommendation framework}~\cite{yin2024device} that can continually adapt to evolving behaviors while retaining prior preferences, and doing so under resource and privacy constraints.

Federated Learning (FL)~\cite{mcmahan2017communication,yang2019federated,kairouz2021advances} enables decentralized model training without exposing raw data, aligning with user-centric recommendation goals. However, most existing methods~\cite{zhang2024federated,wang2024horizontal,zhang2025multifaceted} assume static user behavior and rely on fixed data snapshots, overlooking the continuous evolution of user preferences and limiting long-term personalization. To address this, Continual Learning (CL)~\cite{de2021continual,wang2024comprehensive} is needed to support incremental model updates while preserving previously acquired knowledge. Despite its importance, how to achieve continual personalization within federated recommendation remains an open question.

However, developing continual recommendations in federated settings presents two fundamental and tightly intertwined challenges. First, \textbf{temporal forgetting} arises as user behaviors evolve and local models are incrementally updated with limited storage and replay. Without access to complete interaction histories, each client gradually loses historical preference signals, weakening its ability to retain evolving user interests. Second, \textbf{collaborative personalization} becomes challenging because recommendation relies on learning from user relationships, yet the large differences in client data make it hard to combine this information without weakening each user’s personalized model. These two challenges are mutually coupled and exacerbate each other, ultimately impairing the system's ability to adapt to users' dynamic interest changes.

To fill the gap, this paper presents a novel \textbf{F}ederated \textbf{C}ontinual framework for \textbf{U}ser-\textbf{C}entric \textbf{R}ecommendation (\textbf{\model}). In the framework, each user acts as a client to learn the local recommendation model, and a centralized server orchestrates the system's collaboration to transmit common knowledge. To solve the temporal forgetting challenge on the client side, we develop a \textit{\textbf{time-aware self-distillation strategy}} that encourages the local model to align with its own historical models. This implicit temporal regularization enables the model to preserve long-term user interests while continuously adapting to evolving behaviors, without relying on external memory or replay buffers. Furthermore, we design an \textbf{\textit{inter-user prototype transfer mechanism}} that leverages similar prototypes from a shared latent space to enrich local user representations. Integrated with client-specific decision logic, this mechanism simultaneously fosters more tailored recommendations and enhances collaborative personalization. We summarize the main contributions of this paper as follows:
\begin{itemize}[topsep=2pt, partopsep=0pt, itemsep=1pt, parsep=0pt]
    \item We present a novel federated framework tailored for continuous recommendation, capturing evolving user preferences while preserving privacy. It lays the groundwork for building user-centric personalization systems.
    \item We develop a time-aware self-distillation strategy to mitigate temporal forgetting and preserve long-term user interests. Additionally, we design an inter-user prototype transfer mechanism that bridges knowledge across users, enabling effective collaborative personalization.
    \item Extensive experiments on four benchmark datasets demonstrate the superior performance of our model compared to the baselines. Further analysis confirms its compatibility and practical applicability.
\end{itemize}

\section{Related Work}
\subsection{Continual Recommendation}

In real-world applications, users' interests change over time due to various factors such as seasonal trends or shifting personal tastes. To address this limitation, continual recommendation aims to adapt to user behavior shifts~\cite{yoo2025continual} and learn from new interactions while preserving past knowledge~\cite{cai2022reloop,lee2024continual}. By incorporating strategies in continual learning, such as replay-based approaches~\cite{experience_replay1,experience_replay2} and regularization-based approaches~\cite{learningwithoutforgetting,variationalcl}, continual recommender systems can maintain long-term user preferences while adapting to emerging trends, making them suitable for real-world deployment. These methods have demonstrated effectiveness in sequential~\cite{lee2025leveraging,liu2023linrec} and session-based~\cite{mi2020ader} recommendation scenarios. However, existing methods are typically designed under cloud-based settings and rely on storing all user data. Such centralized paradigms overlook the growing need for user-centric solutions and pose serious user privacy leakage risks.

\subsection{Federated Recommender System}
Federated recommender systems~\cite{sun2024survey,wang2024horizontal,zhang2024federated,zhang2025personalized} are designed to train models in a distributed manner without collecting users' private data, which has been widely recognized as an effective approach to preserving user privacy while enabling personalized recommendations. Early works like MetaMF~\cite{MetaMF} and FedMF~\cite{FedMF} focus on adapting traditional recommendation models to the federated setting, while recent methods such as PFedRec~\cite{PFedRec} and FedRAP~\cite{FedRAP} incorporate client-specific adaptations to enhance personalization modeling. Graph-based methods like FedPerGNN~\cite{FedPerGNN} and FedSage~\cite{FedSage} further enable cross-client collaboration via subgraph expansion and aggregation. Despite the progress in federated recommendation, most existing approaches still rely on static user data and lack the ability to adapt to users’ evolving preferences over time. Recent work F\textsuperscript{3}CRC~\cite{lim2025federated} concentrates on mitigating temporal forgetting under incremental data updates, while neglecting both the streaming nature inherent to practical session-based recommendations and the phenomenon of spatial forgetting across heterogeneous clients. In this paper, we propose a novel federated continuous recommendation framework that captures both temporal continuity and inter-client collaboration. Our method leverages each client’s historical behavioral patterns while enabling knowledge sharing across users, thereby enhancing the personalization and adaptability of recommendations in dynamic environments.

\section{Preliminary}
This section provides a formal description of the user-centric federated recommendation framework. Let $\mathcal{U}$ represent the set of users in the recommender system, with $N=|\mathcal{U}|$ denoting the total number of users. Under the federated learning framework, each user $u \in \mathcal{U}$ operates as a client and learns a recommendation model $\mathcal{F}_\theta$ with personal interaction data $\mathcal{R}_u$ preserved on their own device. 

The training proceeds in multiple iterative rounds between the centralized server and distributed clients, where each user locally updates the model with personal data and uploads the parameters to the server for aggregation into a global model, which is then redistributed for the next round. We formulate the training procedure into the following optimization objective,
\begin{equation}\label{eq:ob_FR}
    \min_{\theta} \sum_{u=1}^N \alpha_{u} \mathcal{L}_{rec}(\theta; \mathcal{R}_u)
\end{equation}
where $\mathcal{L}_{rec}$ is the local recommendation loss function, and $\alpha_u$ is the $u$-th client's aggregation weight, typically set in proportion to client data size relative to the total system data~\cite{mcmahan2017communication}.

\section{Methodology}
\begin{figure*}[t]
    \setlength{\abovecaptionskip}{1mm}
    \setlength{\belowcaptionskip}{-3mm}
    \centering
    \includegraphics[width=1.0\linewidth]{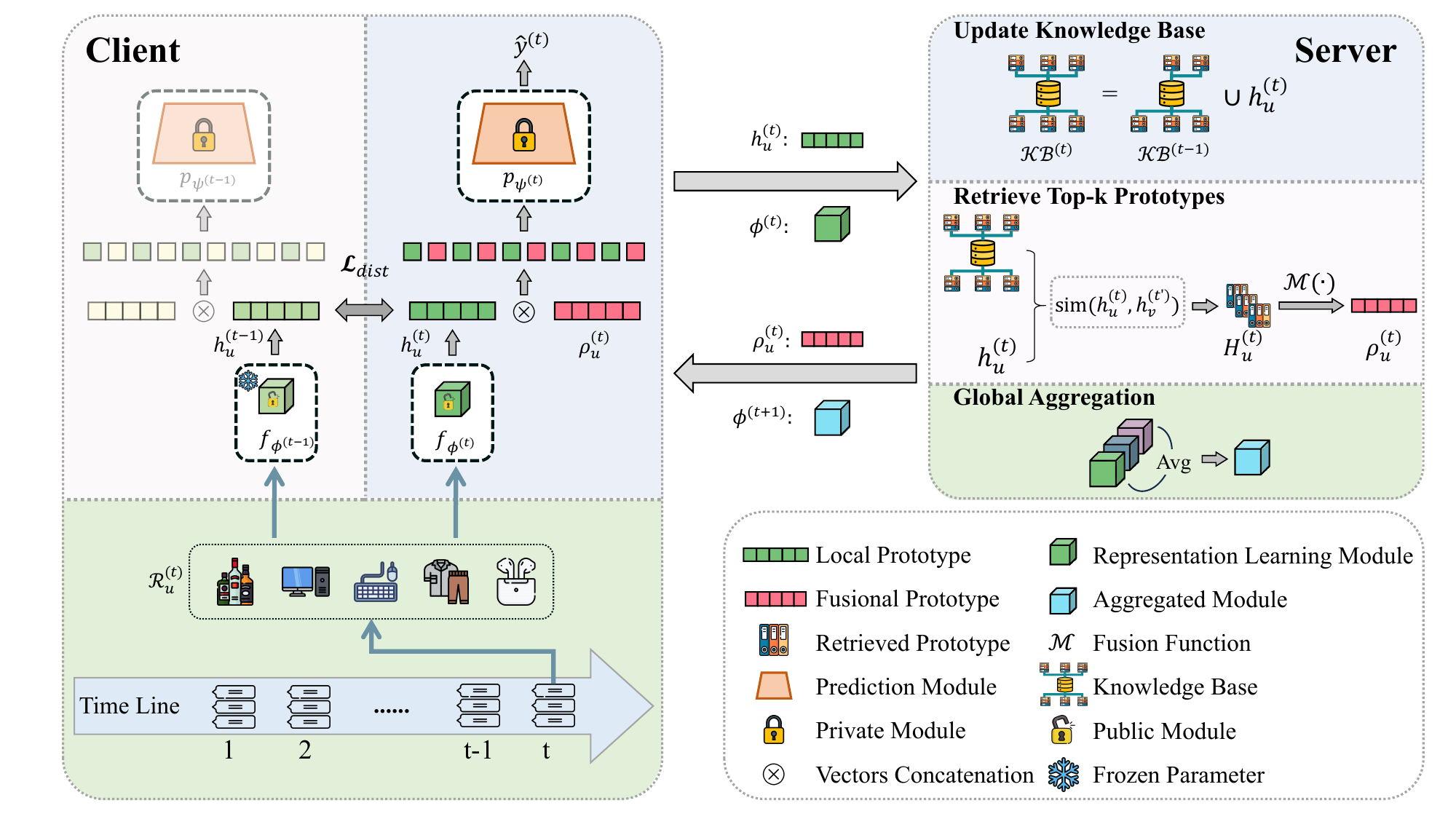}
    \caption{Overall framework of our proposed \model. We focus on continual recommendation, where clients receive new interaction sessions over time. To mitigate temporal forgetting, we apply a time-aware self-distillation strategy that uses the representation module from the previous time step to guide the current session’s learning. Additionally, We introduce an inter-user prototype transfer mechanism where the server maintains a dynamic prototype knowledge base and retrieves the top-$k$ similar prototypes per client to enable collaborative personalization. The figure illustrates the procedure at time step $t$.}
    \label{pipeline}
\end{figure*}

In this section, we introduce the proposed \textbf{F}ederated \textbf{C}ontinual framework for \textbf{U}ser-\textbf{C}entric \textbf{R}ecommendation, abbreviated for \textbf{\model}. Building upon the general user-centric federated recommendation framework, we further present a \textbf{\textit{time-aware self-distillation strategy}} and a \textbf{\textit{inter-user prototype transfer mechanism}} to solve the temporal forgetting and enhance collaborative personalization modeling. This framework empowers personalized recommendations to evolve with user behaviors over time, while maintaining historical knowledge and adhering to privacy constraints. The overall framework is illustrated in Figure~\ref{pipeline}, and next we will detail the framework's components.

\subsection{Time-Aware Self-Distillation Strategy}
In continual recommendation scenarios, a user's personal data accumulates over time, \ie $\mathcal{R}_u=\{\mathcal{R}_u^{(t)}\}_{t=1}^T$, where new interactions are continuously added. Storing the entire history poses challenges for resource-constrained devices, making it impractical to retain all past data. As a result, recommendation models must be continuously updated to adapt to evolving user behaviors. However, training solely on new data often leads to catastrophic forgetting, where valuable historical preferences are overwritten, impairing the model’s ability to capture long-term interest dynamics.

The recommendation model $\mathcal{F}_\theta$ is commonly structured with two components, \ie a representation learning module $f_{\phi}$ that transforms items into latent representations, and a prediction module $p_{\psi}$ that makes recommendations on the representations. In this paper, we focus on sequential recommendation models, where a user's current interactions are influenced by their past behaviors. In this setting, the representation learning module $f_{\phi}$ is typically implemented using a Transformer to capture temporal dependencies, while the prediction module $p_{\psi}$ is often realized with a multilayer perceptron (MLP). Given an input item sequence at time step $t$, denoted as $i_1, i_2, \dots, i_{m-1}$, where $m = |\mathcal{R}^{(t)}|$, the recommendation model predicts the next multiple items $i_2, \dots, i_m$. The overall computation can be formalized as,
\begin{equation}\label{eq:recommendation}
    \hat{\mathbf{y}}^{(t)} = \mathcal{F}_\theta(\mathcal{R}^{(t)}) = p_\psi(f_\phi(i_1, i_2, \dots, i_{m-1}))
\end{equation}
where $\hat{\mathbf{y}}^{(t)}$ denotes the prediction results over the item set for recommendation at time $t$.

In continuous recommendation scenarios, user interactions may occur across different time periods. However, users tend to retain stable cognitive interpretations of item semantics, indicating a consistent underlying perception over time. The representation module $f_{\phi}$ captures users’ semantic understanding of items. To maintain this stable perception over time, we propose a time-aware self-distillation strategy to address catastrophic forgetting. At each time step $t$, we regularize $f_{\phi}^{(t)}$ using its previous state $f_{\phi}^{(t-1)}$ to ensure continuity in user semantics. This is implemented by introducing a self-distillation loss alongside the recommendation loss, formally defined as follows,
\begin{equation}\label{eq:ob_dist}
    \mathcal{L}_{dist} = \text{MSE}(f_{\phi}^{(t)}(\mathcal{R}^{(t)}), f_{\phi}^{(t-1)}(\mathcal{R}^{(t)}))
\end{equation}
where MSE is the mean square error that calculates the distance between the latent representations data $\mathcal{R}^{(t)}$ derived from the current representation module $f_{\phi}^{(t)}$ and the previous one $f_{\phi}^{(t-1)}$. This strategy enables the model to inherit users’ semantic understanding of items without the need to store complete historical data. It strikes a balance between efficiency and continuity, ensuring long-term preference retention under limited storage constraints.

\subsection{Inter-User Prototype Transfer Mechanism}
In federated settings, common knowledge is shared by aggregating model parameters from distributed clients. However, due to significant heterogeneity across local data distributions, direct aggregation tends to be dominated by a few clients and fails to capture meaningful cross-client relationships. This limitation is especially critical for recommendation tasks, where personalized preferences and collaborative patterns are closely intertwined. Therefore, it is essential to design collaborative personalization mechanisms that balance global generalization and local adaptation. 

To enable collaborative personalization, we first reconsider the client recommendation model structure. The representation learning module $f_{\phi}$ extracts item semantic features, which tend to be generalizable across clients, while the prediction module $p_{\psi}$ captures individual behavioral tendencies that vary significantly. Therefore, we propose sharing only the representation learning module $f_{\phi}$ among clients, while keeping each client’s prediction module $p_{\psi}$ private to preserve personalized decision-making. 

Furthermore, we observe that users often exhibit recurring temporal intent patterns, such as transitions from exploration to decision-making and stable behavioral rhythms over time, which reflect shared behavioral structures beyond individual clients. To leverage these regularities, we introduce an inter-user prototype transfer mechanism that enables each client to retrieve semantically aligned intent cues from others and improve behavior modeling with limited data. 

Specifically, a dynamic knowledge base ($\mathcal{KB}$) is maintained on the server. At each time step $t$, after completing local updates, client $u$ uploads both the shared parameters $\phi_u$ and the prototype produced by the local representation learning module $f_{\phi}$ over $\mathcal{R}_u^{(t)}$, denoted as $h_u^{(t)} = f_\phi(\mathcal{R}_u^{(t)})$. The knowledge base is thus updated as follows,
\begin{equation}\label{eq:KB}
    \mathcal{KB}^{(t)} = \mathcal{KB}^{(t-1)} \cup \{h_u^{(t)}\}_{u=1}^N
\end{equation}

Based on the knowledge base, the server retrieves the top-$k$ semantically aligned prototypes for each client $u$, according to the following calculation process,
\begin{equation}\label{eq:similarity}
    H_u^{(t)} = \text{Top-}k\left( \left\{ \text{sim}\left(h_u^{(t)}, h_v^{(t')}\right) \mid h_v^{(t')} \in \mathcal{KB}^{(t)},\ v \ne u \right\} \right)
\end{equation}
where $\text{sim}(\cdot,\cdot)$ denotes a similarity function (\eg cosine similarity), and $H_u^{(t)}$ denotes the set of $k$ selected prototypes most similar to $h_u^{(t)}$. After obtaining $H_u^{(t)}$, the server performs a fusion procedure to derive a consolidated prototype,
\begin{equation}
\label{eq:fusion_H}
   \rho_u^{(t)} =\mathcal{M}(H_u^{(t)})
\end{equation}
where $\mathcal{M}(\cdot)$ is the fusion function, which can be implemented using simple averaging or the more expressive dimensionality reduction techniques.

Upon receiving $\rho_u^{(t)}$, client $u$ integrates the fused prototypes with local features to enrich intent representation and refine user preference modeling. Specifically, we rewrite the recommendation procedure in Equation~(\ref{eq:recommendation}) as follows,
\begin{equation}\label{eq:final_rec}
    % \hat{\mathbf{y}}^{(t)} = p_\psi(h_u^{(t)}||\mathcal{M}(H_u^{(t)}))
    \hat{\mathbf{y}}^{(t)} = p_\psi(h_u^{(t)}||\rho_u^{(t)})
\end{equation}
where $||$ denotes the concatenation operation.

% Upon receiving $H_u^{(t)}$, client $u$ integrates the retrieved prototypes with local features to enrich intent representation and refine user preference modeling. Specifically, we rewrite the recommendation procedure in Equation~(\ref{eq:recommendation}) as follows,
% \begin{equation}\label{eq:final_rec}
%     \hat{\mathbf{y}}^{(t)} = p_\psi(h_u^{(t)}||\mathcal{M}(H_u^{(t)}))
% \end{equation}
% where $||$ denotes the concatenation operation and $\mathcal{M}(\cdot)$ is the fusion function that encodes $H_u^{(t)}$ into a compact vector capturing the prototype-guided intent representation. For example, $\mathcal{M}(\cdot)$ can be instantiated as a simple average function, or a more expressive parameterized fusion module such as an MLP or attention mechanism.

\subsection{Optimization Objective and Algorithm}
\textbf{Optimization objective.}
\model is a federated framework tailored for continual recommendation. Following the above procedure, the overall optimization objective in Equation~(\ref{eq:ob_FR}) can be reformulated as follows,
\begin{equation}\label{eq:ob_ours}
    \min_{\phi, \{\psi_u\}_{u=1}^N} \sum_{u=1}^N \alpha_{u} (\mathcal{L}_{rec}(\theta; \mathcal{R}_u) + \lambda \cdot \mathcal{L}_{dist})
\end{equation}
where $\lambda$ is the coefficient balancing the distillation loss and the main recommendation objective. In this formulation, the representation module parameters $\phi$ are shared across clients, whereas $\psi_u$ captures user-specific decision logic.

\begin{algorithm}[!t]
    \begin{minipage}{1.0\linewidth}
    \caption{Federated Continual Framework for User-Centric Recommendation}
    \label{algo}
    \hspace*{0.05in}{\bf ServerProcedure}:
        \begin{algorithmic}[1]
            \STATE Initialize globally shared representation learning module parameter $\phi^{(1)}$
            \STATE Initialize knowledge base $\mathcal{KB} \gets \emptyset$
            \FOR{each round $t=1, 2, \dots, T$}
                 \FOR{client $u \in \mathcal{U}$ \textbf{in parallel}}
                    \STATE $h_u^{(t)}$ $\leftarrow$ ClientPrototypeCalculation($u, t, \phi^{(t)}$)
                    \STATE Update knowledge base $\mathcal{KB}^{(t)}$ with Eq.~(\ref{eq:KB})
                    \STATE Retrieve top-$k$ semantically aligned prototypes $H_u^{(t)}$ for each client $u$ with Eq.~(\ref{eq:similarity})  
                    \STATE Compute fused prototype $\rho_u^{(t)}$ for each client $u$ with Eq.~(\ref{eq:fusion_H})
                 \ENDFOR
                 \FOR{client $u \in \mathcal{U}$ \textbf{in parallel}}
                     \STATE $\phi_u^{(t)}$ $\leftarrow$ ClientUpdate($u, t, \rho_u^{(t)}$)
                 \ENDFOR
            \STATE $\phi^{(t+1)}$ $\leftarrow$ \text{GlobalAgg}($\phi_u^{(t)}$)
            \ENDFOR
        \end{algorithmic}   
    \hspace*{0.05in}\textbf{ClientPrototypeCalculation}($u, t, \phi^{(t)}$):
        \begin{algorithmic}[1]
            \STATE Initialize $\phi$ with $\phi^{(t)}$
            \STATE Calculate prototype at time step $t$ with $h^{(t)} = f_\phi(\mathcal{R}_u^{(t)})$
            \STATE \textbf{Return} the $t$-th time step prototype $h^{(t)}$
        \end{algorithmic}
    \hspace*{0.05in}\textbf{ClientUpdate}($u, t, \rho_u^{(t)}$):
        \begin{algorithmic}[1]
            \IF{$t = 1$}
                \STATE Initialize prediction module parameter $\psi$
            \ELSE
                \STATE Load $\psi$ from the last round
            \ENDIF
            \FOR{$e = 1, 2, \dots, E$}
                \STATE Perform prediction on $\mathcal{R}_u^{(t)}$ at time step $t$ using the prototype-enhanced formulation in Eq.~(\ref{eq:final_rec})
                \STATE Update $\phi, \psi$ with recommendation loss $\mathcal{L}_{rec}$ and distillation loss $\mathcal{L}_{dist}$
            \ENDFOR
            \STATE \textbf{Return} representation learning module parameter $\phi$
        \end{algorithmic}
    \end{minipage}
\end{algorithm}
\noindent \textbf{Algorithm.}
To solve the optimization objective in Equation~(\ref{eq:ob_ours}), we conduct a federated continual learning algorithm that iteratively coordinates model updates between the server and clients. As illustrated in Algorithm~\ref{algo}, \textbf{before the collaborative training begins}, the server initializes the globally shared representation learning module parameter $\phi$ and sets the knowledge base $\mathcal{KB}$ to an empty set (\textit{lines 1-2 in ServerProcedure}).

\textbf{In each communication round $\bm{t}$}, the client $u \in \mathcal{U}$ first computes its local prototype $h_u^{(t)}$ using the current shared representation module $\phi^{(t)}$ and its personal interaction data $\mathcal{R}_u^{(t)}$. These prototypes are uploaded to the server to update the knowledge base. The server then retrieves the top-$k$ semantically aligned prototypes $H_u^{(t)}$ for each client (\textit{lines 4–7 in ServerProcedure}). Upon receiving the retrieved prototypes, the client updates its local model by incorporating prototype-enhanced representations for recommendation prediction. The local training jointly optimizes the shared module $\phi$ and the personalized module $\psi$ using both the recommendation loss $\mathcal{L}_{rec}$ and distillation loss $\mathcal{L}_{dist}$ (\textit{in ClientUpdate}). After training, the updated shared parameters $\phi$ are sent to the server, where the global update $\phi^{(t+1)}$ is computed using federated averaging for the next communication round (\textit{line 13 in ServerProcedure}).

\subsection{Discussion on Practical Adaptability}
\textbf{Deployment Versatility.}
Our framework adopts a modular design that separates recommendation model into representation learning and personalized prediction. This decoupling enables flexible integration with various model backbones, such as different Transformer-based encoders used in our experiments. The temporal forgetting mitigation and collaborative personalization mechanisms are implemented as independent modules, making them easily plug-and-play across different federated or continual recommendation setups. Such modularity ensures adaptability to diverse deployment environments with minimal changes to the underlying architecture.

\noindent \textbf{Efficiency and Privacy Considerations.}
The proposed framework is well-positioned for practical deployment, with inherent support for efficiency improvement and privacy protection. The server-side prototype knowledge base facilitates collaborative personalization modeling without requiring the exchange of raw data or full model parameters. To further adapt to real-world deployment environments, the framework allows additional refinement such as constraining the size of the knowledge base and reducing prototype transfer frequency. Meanwhile, it is compatible with mainstream privacy-preserving techniques such as differential privacy and secure aggregation, making it extensible to scenarios with stricter regulatory or infrastructural constraints.

\section{Experiment}
\subsection{Experimental Setup}
% dataset, dataset splitting, evaluation metric, baseline, implementation details.
\textbf{Dataset.} We evaluate our method on four real-world benchmark datasets: XING\footnote{\url{https://www.recsyschallenge.com/2017/\#dataset}}~\cite{abel2017recsys}, RetailRocket\footnote{\url{https://www.kaggle.com/datasets/retailrocket/ecommerce-dataset}},  Tmall\footnote{\url{https://tianchi.aliyun.com/dataset/42}}, and LastFM\footnote{\url{http://ocelma.net/MusicRecommendationDataset/lastfm-1K.html}}.

XING and Tmall contain anonymized user-job and user-shopping interactions, respectively. RetailRocket includes six months of e-commerce browsing logs from a Kaggle competition, while LastFM records detailed user listening behavior for music recommendation. For each dataset, we segment users' interaction histories into multiple sessions based on timestamps to simulate a sequential recommendation setting, following the preprocessing procedure of HRNN~\cite{HRNN} to ensure consistency and fair comparison. Taking the XING dataset as an example, user interactions are segmented into sessions based on a 30-minute inactivity threshold, and interactions labeled as `DELETE' are removed. The test set is constructed from each user’s last session, while the second-to-last session forms the validation set. The remaining sessions comprise the training set. Additionally, items in the test and validation sets that do not appear in the training set are filtered out. All datasets are preprocessed by dividing continuous interaction sequences into sessions using an inactivity threshold, following the standard protocol in session-based recommendation~\cite{HRNN,DMI-GNN,SASRec}. Each session thus represents a temporally coherent sequence of user interactions, enabling the construction of session-level recommendation tasks. The statistics of the processed datasets are summarized in Table~\ref{tab:dataset_statistics}.

 \begin{table}[!t]
\centering
\caption{Datasets statistics.}
\begin{tabular}{p{50pt}|p{25pt}<{\centering}|p{30pt}<{\centering}|p{30pt}<{\centering}|p{40pt}<{\centering}}
\toprule
\multirow{2}{*}{\textbf{Dataset}} & \multirow{2}{*}{\textbf{Users}} & \multirow{2}{*}{\textbf{Items}} & \multirow{2}{*}{\textbf{Sessions}} & \textbf{Averaged} \\
& & & & \textbf{Length} \\
\midrule
XING           & 877  & 29,366  & 6,165 & 15.10 \\
RetailRocket   & 391  & 14,051  & 4,183 & 7.93 \\
LastFM         & 940  & 145,469  & 468,297 & 18.50 \\
Tmall          & 1,746  & 123,405  & 12,594 & 37.78 \\
\bottomrule
\end{tabular}

\label{tab:dataset_statistics}
\end{table}

% More details about dataset descriptions, preprocessing procedure and statistics are summarized in the Appendix.

\noindent \textbf{Baselines.} Since there are no existing methods tailored for continual recommendation under the federated setting, we select three state-of-the-art federated recommendation models as baselines. Additionally, we adapt two categories of relevant centralized recommendation approaches to the federated setting, in order to enrich the baseline set and provide a broader evaluation perspective. To distinguish these adapted methods, we prepend `Fed-' to their names in the experimental tables. The specific methods in each category are as follows:

% \textbf{(1) Federated recommendation methods:} PFedRec~\cite{PFedRec}, FedRAP~\cite{FedRAP}, and GPFedRec~\cite{gpfedrec}.

% \textbf{(2) Recent state-of-the-art session-based methods:} DMI-GNN~\cite{DMI-GNN}, HearInt~\cite{HearInt}, MiaSRec~\cite{MiaSRec}.

% \textbf{(3) Representative sequential methods:} GRU4Rec~\cite{GRU4Rec}, SASRec~\cite{SASRec}, and TiSASRec~\cite{TiSASRec}.

\textbf{(1) Federated recommendation methods:} \textbf{PFedRec}~\cite{PFedRec} incorporates client-specific adaptations to enhance personalization modeling, allowing each client to maintain a personalized model. \textbf{FedRAP}~\cite{FedRAP} learns a global view of items via federated learning while simultaneously maintaining a personalized view locally on each user. \textbf{GPFedRec}~\cite{gpfedrec} constructs a user-relation graph on the server from user-specific personalized item embeddings, enabling effective modeling of cross-user relations and improving recommendation accuracy under heterogeneity.

\textbf{(2) Recent state-of-the-art session-based methods:} \textbf{DMI-GNN}~\cite{DMI-GNN} introduces a multi-interest learning framework into session-based recommendation.The framework is optimized by a multi-position pattern (MPP) learning method and a dynamic multi-interest (DMI) regularization method. \textbf{HearInt}~\cite{HearInt} designs a Temporal Intent Disentanglement (TID) module and a cross-scale contrastive learning task to capture both short-term and long-term user intents, mitigating mutual interference between them and improving intent representation. \textbf{MiaSRec}~\cite{MiaSRec} dynamically constructs multiple session representations centered around each item and selects key features to model user’s diverse intents, enabling intent-aware recommendations in session scenarios.

\textbf{(3) Representative sequential methods:} \textbf{GRU4Rec}~\cite{GRU4Rec} is one of the RNN-based sequential recommendation models. It employs gated recurrent units (GRU) to capture sequential dependencies in user behavior within sessions. \textbf{SASRec}~\cite{SASRec} adopts a self-attention mechanism to model user behavior sequences, enabling the model to learn both short and long range dependencies. \textbf{TiSASRec}~\cite{TiSASRec} extends SASRec by incorporating time intervals and item positions into the attention mechanism, thereby enhancing the model’s ability to capture temporal dynamics and position-aware preferences in user interactions.

%adding to appendix
% The first category consists of traditional sequential and session-based recommendation methods: GRU4Rec~\cite{GRU4Rec}, SASRec~\cite{SASRec}, and TiSASRec~\cite{TiSASRec}.

% The second category includes recent state-of-the-art (SotA) methods in session-based recommendation under centralized settings: DMI-GNN~\cite{DMI-GNN}, HearInt~\cite{HearInt}, MiaSRec~\cite{MiaSRec}.

% The third category encompasses recommendation approaches tailored to federated learning scenarios, aiming to address challenges such as data decentralization and client heterogeneity. Representative methods include PFedRec~\cite{PFedRec}, FedRAP~\cite{FedRAP}, and GPFedRec~\cite{gpfedrec}.

% 这里具体的细节加到附录
% In particular, for graph-based models in the second category, each client constructs a local subgraph based solely on their own interactions, in line with privacy constraints and data isolation requirements inherent in federated learning.

\noindent \textbf{Implementation details.}
We adopt two widely used evaluation metrics in recommender systems: Hit Rate (HR@n) and Normalized Discounted Cumulative Gain(NDCG@n)~\cite{DMI-GNN,xie2022contrastive}. We implement our proposed method on three representative sequential recommendation backbones. All analytical experiments are conducted on dataset XING using SASRec as the unified backbone architecture to ensure consistency. More implementation details can be found in Appendix~A.1.

\begin{table*}[!t]
\renewcommand{\arraystretch}{1.2}
\centering
\caption{Overall performance comparison between our proposed \model and baseline methods. Bold numbers indicate the best results, underlined values represent the second-best, and ``*'' marks statistically significant improvements over the backbone baseline (two-sided t-test, $p<0.05$).}
\resizebox{\textwidth}{!}{
\begin{tabular}{p{60pt}<{\centering}|p{70pt}<{\centering}|p{40pt}<{\centering}p{45pt}<{\centering}|p{40pt}<{\centering}p{45pt}<{\centering}|p{40pt}<{\centering}p{45pt}<{\centering}|p{40pt}<{\centering}p{45pt}<{\centering}}
\toprule[1.0pt]
\multirow{2}{*}{\textbf{Type}} & \multirow{2}{*}{\textbf{Model}} & \multicolumn{2}{c|}{\textbf{XING}} & \multicolumn{2}{c|}{\textbf{RetailRocket}} & \multicolumn{2}{c|}{\textbf{LastFM}} & \multicolumn{2}{c}{\textbf{Tmall}} \\
& & HR@10 & NDCG@10 & HR@10 & NDCG@10 & HR@10 & NDCG@10 & HR@10 & NDCG@10\\
\midrule[0.2pt]
\multirow{3}{*}{\textbf{Federated}}&\textbf{PFedRec} & 0.1852 & 0.0898 & 0.2603 & 0.1706 & 0.1596 & 0.0835 & 0.2669 & 0.1451 \\
&\textbf{FedRAP} & 0.2013 & 0.1318 & 0.3183 & 0.2198 & 0.2191 & 0.1198 & 0.2411 & 0.1309 \\
&\textbf{GPFedRec} & 0.1804 & 0.0839 & 0.2303 & 0.1322 & 0.1532 & 0.0777 & 0.2216 & 0.1208
 \\
\midrule[0.2pt]
\multirow{3}{*}{\textbf{Session-based}} & \textbf{Fed-DMI-GNN} & 0.2561 & 0.1584 & 0.2876 & 0.1983 & 0.3216 & \underline{0.2060} & 0.2068 & 0.1145 \\
& \textbf{Fed-HearInt} & 0.1387 & 0.0708 & 0.2305 & 0.1293 & 0.2404 & 0.1302 & 0.2027 & 0.1182 \\
& \textbf{Fed-MiaSRec} & 0.1692 & 0.0910 & 0.2473 & 0.1439 & 0.2894 & 0.1756 & 0.2451 & 0.1385 \\
\midrule[0.2pt]
\multirow{6}{*}{\textbf{Sequential}} &\textbf{Fed-GRU4REC} & 0.1765 & 0.0939 & 0.2392 & 0.1432 & 0.3291 & 0.1918 & 0.2274 & 0.1248 \\
&\textbf{w/ \model} & 0.2311* & 0.1563* & 0.3454* & 0.2687* & 0.3362* & 0.1946* & 0.2394* & 0.1420*  \\
\cmidrule(lr){2-10}
&\textbf{Fed-SASRec} & 0.1800 & 0.0954 & 0.2574 & 0.1527 & 0.3271 & 0.1907 & 0.2898 & 0.1571 \\
&\textbf{w/ \model} & \underline{0.2882}* & \underline{0.1863}* & \underline{0.3602}* & \underline{0.2879}* & \underline{0.3426}* & 0.1944* & \textbf{0.3074}* & \textbf{0.1947}* \\
\cmidrule(lr){2-10}
&\textbf{Fed-TiSASRec} & 0.2375 & 0.1549 & 0.2608 & 0.1602 & 0.3359 & 0.1964 & 0.2718 & 0.1507 \\
&\textbf{w/ \model} & \textbf{0.2983}* & \textbf{0.1891}* & \textbf{0.3750}* & \textbf{0.2996}* & \textbf{0.3553}* & \textbf{0.2187}* & 
\underline{0.2967}* & \underline{0.1593}* \\
\bottomrule[1.0pt]
\end{tabular}
}

\label{tab:overall_performance}
\end{table*}

\begin{table*}[!t]
\small
\renewcommand{\arraystretch}{1.2}
\centering
\caption{Ablation results by removing two key components from our proposed \model, denoted as w/o TASD (time-aware self-distillation) and w/o IUPR (inter-user prototype transfer) respectively.}
\resizebox{\textwidth}{!}{
\begin{tabular}{p{50pt}|p{40pt}<{\centering}p{45pt}<{\centering}|p{40pt}<{\centering}p{45pt}<{\centering}|p{40pt}<{\centering}p{45pt}<{\centering}|p{40pt}<{\centering}p{45pt}<{\centering}}
\toprule[1.0pt]
\multirow{2}{*}{\textbf{Model}} & \multicolumn{2}{c|}{\textbf{XING}} & \multicolumn{2}{c|}{\textbf{RetailRocket}} & \multicolumn{2}{c|}{\textbf{LastFM}} & \multicolumn{2}{c}{\textbf{Tmall}} \\
& HR@10 & NDCG@10 & HR@10 & NDCG@10 & HR@10 & NDCG@10 & HR@10 & NDCG@10\\
\midrule[0.2pt]
\textbf{\model} & \textbf{0.2882} & \textbf{0.1863} & \textbf{0.3602} & \textbf{0.2879} & \textbf{0.3426} & \textbf{0.1944} & \textbf{0.3074} & \textbf{0.1947} \\
% \textbf{Ours w/o transfer zero} & 0.3374 & 0.2705 & 0.2580 & 0.1740 & 0.3298 & 0.1907 & 0.2715 & 0.1497 \\
% \textbf{Ours w/o transfer random} & 0.3038 & 0.2392 & 0.2570 & 0.1770 & 0.2989 & 0.1777 & 0.2772 & 0.1506 \\
\textbf{w/o TASD} & 0.2731 & 0.1795 & 0.3360 & 0.2590 & 0.3309 & 0.1967 & 0.2926 & 0.1860 \\
\textbf{w/o IUPR} & 0.2556 & 0.1754 & 0.3324 & 0.2692 & 0.3287 & 0.1912 & 0.2520 & 0.1412\\
\bottomrule[1.0pt]
\end{tabular}
}

\label{tab:Ablation study}
\end{table*}

\subsection{Overall Comparison}
We implement our proposed method on three representative sequential recommendation backbones and compare with three categories of baselines across four benchmark datasets. The results are presented in Table~\ref{tab:overall_performance}. The experimental outcomes suggest the following key points:

\textit{\textbf{(1) Our method outperforms existing federated recommendation methods.}}
While existing federated recommendation methods primarily focus on client-side personalization, they often overlook temporal dependencies within each user. This limits their ability to model evolving preferences, which are crucial in continual recommendation tasks. In contrast, our method integrates a time-aware self-distillation strategy that captures temporal signals from past sessions. This enables the model to preserve sequential patterns over time, thereby enhancing its ability to model user dynamics in federated environments.

\textit{\textbf{(2) Our method outperforms session-based recommendation models.}} Existing session-based methods typically leverage multiple user interests or graph-based modeling to capture local user continual behaviors. However, they lack mechanisms for inter-client collaboration when extended to federated environments, limiting global information sharing. In contrast, our approach incorporates an inter-user prototype transfer mechanism that propagates intent patterns across clients. By leveraging globally consistent behavioral cues beyond local data, each client can refine its representations, resulting in consistently improved performance.

\textit{\textbf{(3) Our method exhibits strong compatibility across diverse backbones.}} The proposed framework is model-agnostic and can be seamlessly integrated with existing sequential recommendation models. Applied to three representative backbones, it consistently improves performance across all datasets, demonstrating broad applicability and robust adaptability to different architectures.

\begin{figure*}[!t]
    \setlength{\abovecaptionskip}{1mm}
    \setlength{\belowcaptionskip}{-3mm}
    \centering
    \includegraphics[width=\linewidth]{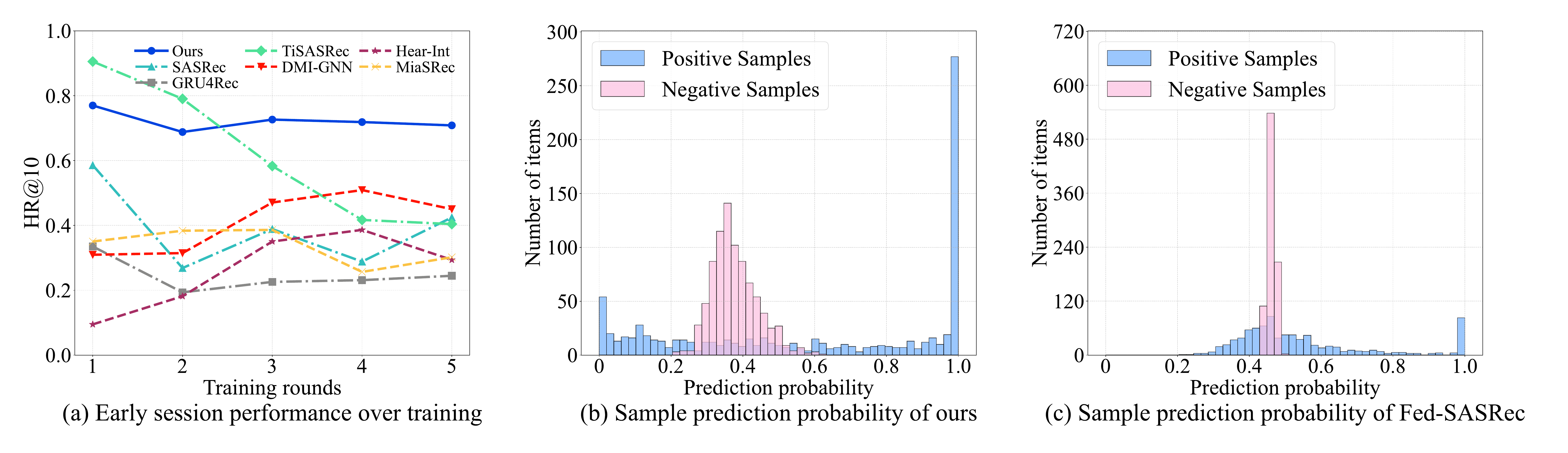}
    \caption{Component effectiveness analysis on XING dataset.}
    \label{Ablation study}
\end{figure*}
\subsection{Component Effectiveness Analysis}
We propose a time-aware self-distillation strategy and an inter-user prototype transfer mechanism to address temporal forgetting and enhance collaborative personalization modeling. To comprehensively validate their contributions, we conduct both quantitative (ablation study) and qualitative (visualization-based) analyses.

\noindent \textbf{Ablation results.} 
To assess the individual contributions of key components in our method, we conduct ablation studies by removing each component from \model. As shown in Table~\ref{tab:Ablation study}, we observe that removing either the time-aware self-distillation strategy or the inter-user prototype transfer mechanism leads to noticeable performance degradation. These results are attributed to the fact that the former helps preserve long-term user interests by encouraging alignment with historical models, while the latter enhances local representations through collaborative personalization by leveraging similar user prototypes. This highlights the complementary roles of the two components in addressing continual recommendations under federated setting.

\noindent \textbf{Visualization-based interpretations.} 
To gain deeper insights into the roles of the key components beyond numerical results, we conduct visualization-based analyses to better evaluate the effectiveness of these two modules.

We propose the \textbf{\textit{time-aware self-distillation strategy}} to address the issue of temporal forgetting in continual client recommendation. To evaluate its effectiveness, we assess the model’s resistance to forgetting by re-evaluating each client’s first session after every communication round. Tracking performance changes on this initial session over time allows us to examine how well the model retains early session preferences throughout training. 

To provide an intuitive comparison, we track the performance trajectories of baseline methods with temporal modeling capabilities and our approach by evaluating the initial session throughout training. As shown in Figure~\ref{Ablation study}(a), most baseline methods suffer from performance degradation in early sessions as new sessions accumulate, indicating a tendency toward temporal forgetting. The only exceptions are the graph-based methods DMI-GNN and HearInt, whose use of graph structures allows them to incorporate information from earlier sessions, sometimes leading to gradual improvement in early-session metrics. In contrast, our method consistently maintains high accuracy across training rounds, demonstrating strong resistance to temporal forgetting.

To enhance collaborative personalization, we design an \textbf{\textit{inter-user prototype transfer mechanism}} that improves behavior modeling by enriching each client’s local features through transferring semantically similar prototypes across clients. Our experiments focus on the implicit recommendation setting, where each interaction is either a positive or negative sample. To evaluate the effectiveness of the prototype transfer, we analyze the distribution of predicted probabilities on test samples after model convergence. For intuitive comparison, we also present results from the baseline Fed-SASRec without prototype transfer. 
% We conduct a visualization experiment by training our method and a SASRec-based model without the prototype mechanism for 20 communication rounds. After training, we compute the average predicted scores for positive and negative samples on the validation set of each client and plot their distributions as histograms.

As shown in Figure~\ref{Ablation study}(b), by leveraging similar prototypes to enhance local features, our method achieves a clear separation between positive and negative samples: most positive samples receive higher prediction probability, and the class distributions become more distinct. In contrast, Figure~\ref{Ablation study}(c) shows that Fed-SASRec without prototype enhancement produces fewer high-confidence positive predictions, with negative probabilities largely clustered around 0.5, reflecting weaker discriminative power. These results demonstrate that the proposed mechanism effectively widens the prediction probability gap between positive and negative samples, resulting in improved recommendation performance.

% To analyze the impact of cross-client knowledge sharing, we design ablation studies by replacing the auxiliary vector with three alternative settings: (1) a zero vector, (2) a randomly initialized vector, and (3) complete removal of the auxiliary vector. These variants help assess the effectiveness of the proposed auxiliary mechanism in facilitating inter-client knowledge transfer. All three settings lead to a noticeable decline in both Hit Rate and ranking performance of \model across all four datasets, indicating that the auxiliary vector plays a crucial role in enhancing recommendation quality.

% A similar decline in both metrics is observed when the historical distillation module is removed, likely because continual updates on new session data cause the model to forget knowledge learned from earlier interactions.

% These ablation results collectively validate the effectiveness of the key components in the \model framework. Both the cross-client knowledge sharing via auxiliary vectors and the local historical model distillation play crucial roles in enhancing recommendation accuracy and stability.

\subsection{Hyper-Parameter Analysis}
This subsection investigates the impact of two key hyperparameters on model performance. 
% For \textbf{the number of semantically aligned prototypes $k$}, we find that model performance peaks with a moderate $k$, which offers a good trade-off between auxiliary information sufficiency and prototype relevance. It avoids both insufficient guidance from too few prototypes and diluted relevance from too many. For \textbf{the coefficient of distillation loss $\lambda$}, we find that model performance achieves the best with a moderate $\lambda$, which effectively balances adapting to new data and retaining historical knowledge. 

\begin{figure}[t]
\setlength{\abovecaptionskip}{1mm}
    \setlength{\belowcaptionskip}{-3mm}
    \centering
    
    \includegraphics[width=1.0\linewidth]{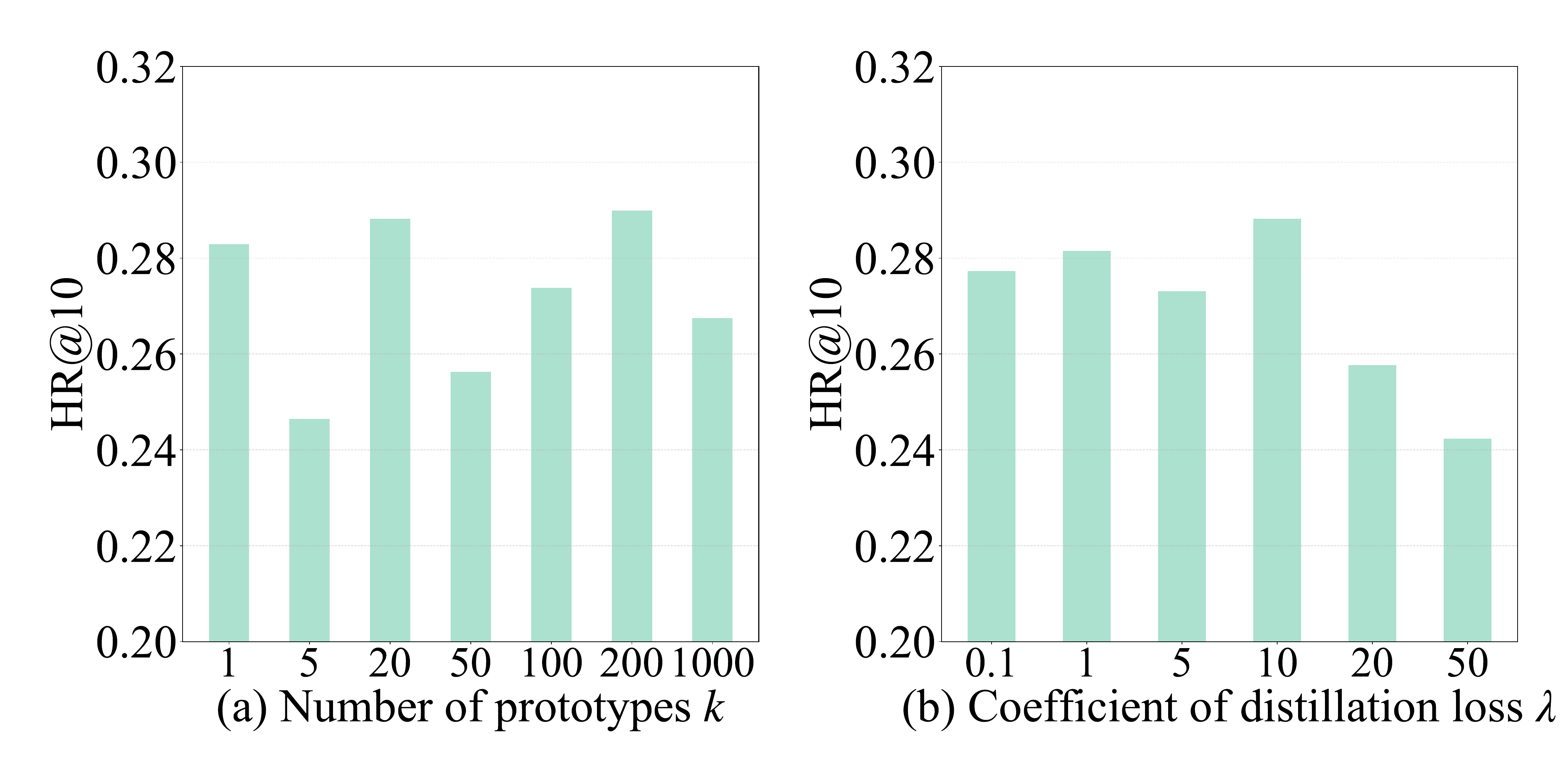}
    \caption{Effect of two hyper-parameters on model performance (HR@10) on the XING dataset.}
    \label{hyperparameter_HR_xing}
\end{figure}

\noindent \textbf{The number of semantically aligned prototypes $k$.} 
To comprehensively evaluate its impact on model performance, we set the values of $k$ among $\{1, 5, 20, 100, 200, 1000\}$. As shown in Figure~\ref{hyperparameter_HR_xing}(a), model performance initially improves as $k$ increases, but starts to decline when $k$ becomes too large. This trend suggests that a small $k$ may lead to insufficient auxiliary information, limiting the benefit of prototype transfer. In contrast, overly large $k$ values can introduce noisy or less relevant prototypes, which may impair model learning. Therefore, a moderate number of transferred prototypes helps strike a better balance between diversity and consistency in collaborative personalization modeling. 
Full experimental results and detailed analysis for all datasets are provided in the Appendix~C.

\noindent \textbf{The coefficient of distillation loss $\lambda$.} 
We set $\lambda$ in \{0.1, 1, 5, 10, 20, 50\} to assess its impact on performance. As shown in Figure~\ref{hyperparameter_HR_xing}(b), the best results are achieved when $\lambda = 10$. A smaller $\lambda$ emphasizes adaptability to new data, while a larger $\lambda$ favors stability by preserving historical knowledge. This confirms that moderate temporal regularization offers a better balance between plasticity and stability. Full experimental results and detailed analysis for all datasets are provided in the Appendix~C.

\subsection{Practical Viability Analysis}
% Compared to conventional FedAvg-based methods, our proposed \model framework requires each client to both transmit and receive an additional session representation vector from the server. This vector encapsulates information specific to a single session and does not retain any user-identifiable data, thereby significantly mitigating the risk of privacy leakage. Nevertheless, the transmission of such session-level information, along with the server-side storage of the session knowledge base, still presents potential privacy concerns that warrant further investigation. 
% Since only a single vector is transmitted, differential privacy and encryption techniques can be applied to this vector during transmission to further enhance privacy protection. In Figure X, we evaluate the impact of applying differential privacy to the transmitted auxiliary vectors by injecting calibrated noise. Although this privacy-preserving mechanism results in a moderate decline in performance, our proposed \model framework still achieves competitive results. Moreover, the method demonstrates robustness across various noise levels, indicating its practical viability under different privacy budgets.

\noindent \textbf{Efficiency improvement.}
As discussed earlier, our proposed \model can be further refined to improve system efficiency. Here, we integrate two lightweight mechanisms to achieve this goal.

\noindent \textbf{(1) Limiting the size of the knowledge base.}
Instead of retaining all client prototypes across communication rounds, we adopt a sliding-window strategy that retains only the client prototypes from the most recent communication rounds $r$. Once the number of rounds exceeds $r$, client prototypes from earlier rounds are discarded and replaced with those from newer ones. As shown in the Appendix~D.1 Table~\ref{tab:KB_restricted}, reducing the size of the prototype knowledge base maintains stable performance and in some cases even improves it, while significantly lowering computational overhead.

\noindent \textbf{(2) Reducing the frequency of prototype transfer.}
We explore the effect of varying the transfer interval, where prototype transfer is performed once every few rounds. Specifically, we set the interval to $\{1, 2, 3, 4, 5\}$, indicating the number of rounds between each transfer. As shown in Appendix~D.1 Table~\ref{tab:different intervals}, choosing an appropriate interval (\eg 3) maintains comparable performance while significantly reducing computational overhead (measured as the total cost within each interval divided by the number of rounds it spans). This demonstrates that less frequent transfers can effectively balance recommendation quality and system efficiency.
% Results on other datasets are reported in the Appendix~D.1.

\subsection{Privacy considerations.}

\model is inherently privacy-friendly, as each client transmits a single session-level representation and representation module parameters without user-identifiable information. This design allows easy integration with privacy-preserving techniques. We apply local differential privacy by injecting Laplacian noise into client parameters, using varying noise scales $std=\{0.1,\ldots,0.6\}$. These correspond to approximately $(\epsilon, \delta)$-LDP with $\delta = 10^{-5}$ and $\epsilon$ ranging from 48.5 (weaker privacy) to 8.1 (stronger privacy), as calculated by the Gaussian mechanism. As shown in Appendix~D.2, \model remains robust under different noise levels, maintaining competitive performance while enhancing privacy protection.
% As shown in Figure~\ref{privacy_xing}, \model remains robust under different noise levels, maintaining competitive performance while enhancing privacy protection.

% In this section, we present the main results on the XING dataset, whereas the results on the other datasets are provided in Appendix~D.2 to ensure the generality and consistency of our findings across different data domains.

% \begin{figure}[t]
%     \setlength{\abovecaptionskip}{1mm}
%     \setlength{\belowcaptionskip}{-3mm}
%     \centering
%     \includegraphics[width=0.8\linewidth]{figures/privacy_xing.pdf}
%     \caption{Performance of integrating LDP into our \model.}
%     \label{privacy_xing}
% \end{figure}

\section{Conclusion}
This paper introduces \model, a federated continual recommendation framework that supports long-term personalization in a privacy-preserving manner. Unlike existing methods that rely on static data and overlook the evolving nature of user behavior, \model is explicitly designed to address the dual challenges of temporal forgetting and collaborative personalization in federated settings. By introducing a time-aware self-distillation strategy, the framework enables each client to retain historical preferences without external memory or data replay. Furthermore, the inter-user prototype transfer mechanism allows knowledge sharing across clients while preserving individual decision logic. Extensive experiments on four public datasets demonstrate that \model consistently outperforms existing baselines, providing a robust, efficient, and privacy-aware solution for dynamic user-centric recommendation.

\begin{acks}
Chunxu Zhang, Zhiheng Xue, Weipeng Zhang and Bo Yang are supported by the National Natural Science Foundation of China under Grant Nos.U22A2098, 62172185, 62206105 and 62202200; the Major Science and Technology Development Plan of Jilin Province under Grant No.20240212003GX, the Major Science and Technology Development Plan of Changchun under Grant No.2024WX05.
\end{acks}

%%
%% The next two lines define the bibliography style to be used, and
%% the bibliography file.
\newpage
\bibliographystyle{ACM-Reference-Format}
\balance
\bibliography{www2026}

%%
%% If your work has an appendix, this is the place to put it.
\appendix

\section{Experimental Setup}

\subsection{More Details about Implementation}
We adopt two widely used evaluation metrics in recommender systems: Hit Rate (HR@n) and Normalized Discounted Cumulative Gain(NDCG@n)~\cite{DMI-GNN,xie2022contrastive}. These metrics assess both the accuracy and the ranking quality of the recommendations, reflecting how well the model can surface target items among the top-n positions. In this work, we fix n=10 for evaluation across all experiments. 
We implement our proposed method on three representative sequential recommendation backbones. To ensure consistency in the analytical experiments, SASRec is adopted as the unified backbone architecture and evaluated on XING dataset. We adapt two categories of representative centralized recommendation approaches to a federated setting by applying the standard FedAvg algorithm, ensuring a fair comparison under distributed learning conditions. For graph-based models in the second category, each client constructs a local subgraph using only its own interaction data, adhering to the privacy constraints and data isolation principles inherent in FL.

The model involves several key hyperparameters, including the learning rate, embedding size, number of local training epochs, total communication rounds, etc. We tune key hyperparameters such as the learning rate and embedding size based on validation performance to ensure optimal model behavior. Specifically, we use the Adam optimizer with a learning rate of 0.1 and set the embedding size to 50. The number of local training epochs is fixed to 4 per communication round and the total number of communication rounds is limited to 40, with an early stopping strategy triggered if no improvement is observed for 5 consecutive rounds. Empirical results show that the model converges reliably under these settings.

Experiments are conducted using six NVIDIA A800 GPUs and three NVIDIA RTX 3090 GPUs with PyTorch 2.4.0 + cu115. All results are averaged over five independent runs to ensure robustness.

\section{Code Availability}

The implementation of our proposed method is publicly available to ensure reproducibility and facilitate further research. All relevant code files and running instructions are accessible at the repository: \url{https://github.com/Poizoner/code4FCUCR_www2026}. Please refer to the README file in the repository for details on environment setup and result reproduction.

% The implementation of our proposed method is publicly available to ensure reproducibility and facilitate further research. All relevant code files can be found in the supplementary material under the directory \texttt{./code4FCUCR}. The execution process, including environment setup and running instructions, is detailed in the provided \texttt{./code4FCUCR/readme.md} file. Please refer to the README for guidance on reproducing our results.

\begin{table*}[!t]
\centering
\caption{Model performance and time cost under various knowledge base sizes on other datasets.}
\begin{tabular}{l|l|c|c|c|c|c}
\toprule
\textbf{Dataset} & \textbf{Metric}  & \textbf{Recent 1 round} & \textbf{Recent 2 round} & \textbf{Recent 3 round} & \textbf{Recent 5 round} & \textbf{All round}\\
\midrule
\multirow{3}{*}{XING}
& HR@10      & 0.2528 & 0.2857 & 0.2801 & 0.2829 & 0.2882\\
& NDCG@10    & 0.1696 & 0.1985 & 0.1949 & 0.1858 & 0.1863\\

& Time per round & 622s & 640s & 649s  & 668s  & 672s\\
\midrule
\multirow{3}{*}{RetailRocket}
& HR@10      & 0.3360 & 0.3427 & 0.3817 & 0.3737 & 0.3602\\
& NDCG@10    & 0.2767 & 0.2744 & 0.2974 & 0.2936 & 0.2879\\

& Time per round & 41s & 44s & 48s  & 50s  & 62s\\
\midrule
\multirow{3}{*}{LastFM}
& HR@10      & 0.2894 & 0.3383 & 0.3596 & 0.3777 & 0.3426\\
& NDCG@10    & 0.1723 & 0.2102 & 0.2213 & 0.2397 & 0.1944\\

& Time per round & 41min & 44min & 48min  & 52min  & 1h8min\\

\midrule
\multirow{3}{*}{Tmall}
& HR@10      & 0.2921 & 0.3144 & 0.2961 & 0.3007 & 0.3074\\
& NDCG@10    & 0.1821 & 0.2034 & 0.1883 & 0.1931 & 0.1947\\
& Time per round & 1h30min & 1h41min & 1h48min  & 1h52min  & 2h13min\\
\bottomrule
\end{tabular}

\label{tab:KB_restricted}
\end{table*}

\begin{table*}[!t]
\centering
\caption{Model performance and time cost under various prototype transfer frequencies on other datasets.}
\begin{tabular}{l|l|c|c|c|c|c}
\toprule
\textbf{Dataset} & \textbf{Metric} & \textbf{Interval=1} & \textbf{Interval=2} & \textbf{Interval=3} & \textbf{Interval=4} & \textbf{Interval=5} \\
\midrule
\multirow{3}{*}{XING}
& HR@10 & 0.2882 & 0.2563 & 0.2689 & 0.2423 & 0.2493 \\
& NDCG@10 & 0.1863 & 0.1786 & 0.1844 & 0.1936 & 0.2031 \\
& Averaged time per round & 672s & 643s & 634s & 619s & 598s \\
\midrule
\multirow{3}{*}{RetailRocket}
& HR@10   & 0.3602 & 0.3562 & 0.3360 & 0.3199 & 0.3118 \\
& NDCG@10 & 0.2879 & 0.2781 & 0.2654 & 0.2648 & 0.2524 \\
& Averaged time per round & 62s & 58s & 54s & 48s & 40s \\
\midrule
\multirow{3}{*}{LastFM}
& HR@10 & 0.3426 & 0.3121 & 0.2867 & 0.2731 & 0.2658 \\
& NDCG@10 & 0.1944 & 0.1890 & 0.1838 & 0.1782 & 0.1825 \\
& Averaged time per round & 1h8min & 1h3min & 0h50min & 0h48min & 0h46min \\
\midrule
\multirow{3}{*}{Tmall}
& HR@10 & 0.3074 & 0.3008 & 0.2965 & 0.2889 & 0.2915 \\
& NDCG@10 & 0.1947 & 0.1921 & 0.1884 & 0.1850 & 0.1873 \\
& Averaged time per round & 2h13min & 1h58min & 1h42min & 1h30min & 1h19min \\
\bottomrule

\end{tabular}

\label{tab:different intervals}
\end{table*}

\section{Hyper-Parameter Analysis}
To complement the main text in Section~5.4, we provide a detailed analysis of the impact of two key hyperparameters $k$ and $\lambda$ on the remaining datasets.

\begin{figure}[t]
    \centering
    \includegraphics[width=1.0\linewidth]{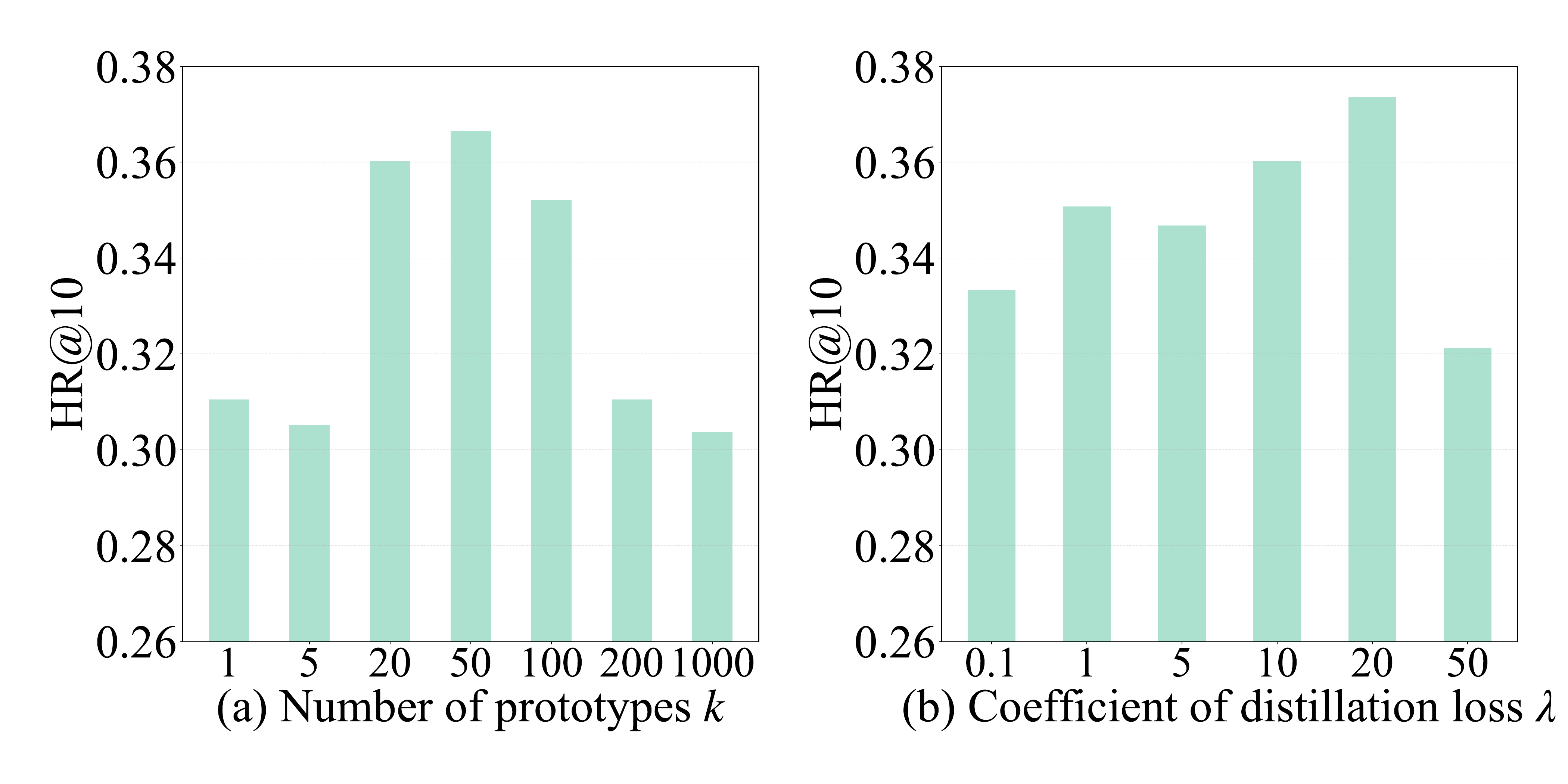}
    \caption{Effect of two hyper-parameters on model performance (HR@10) on the RetailRocket dataset.}
    \label{hyperparameter_HR_retail}
\end{figure}
\begin{figure}[t]
    \centering
    
    \includegraphics[width=1.0\linewidth]{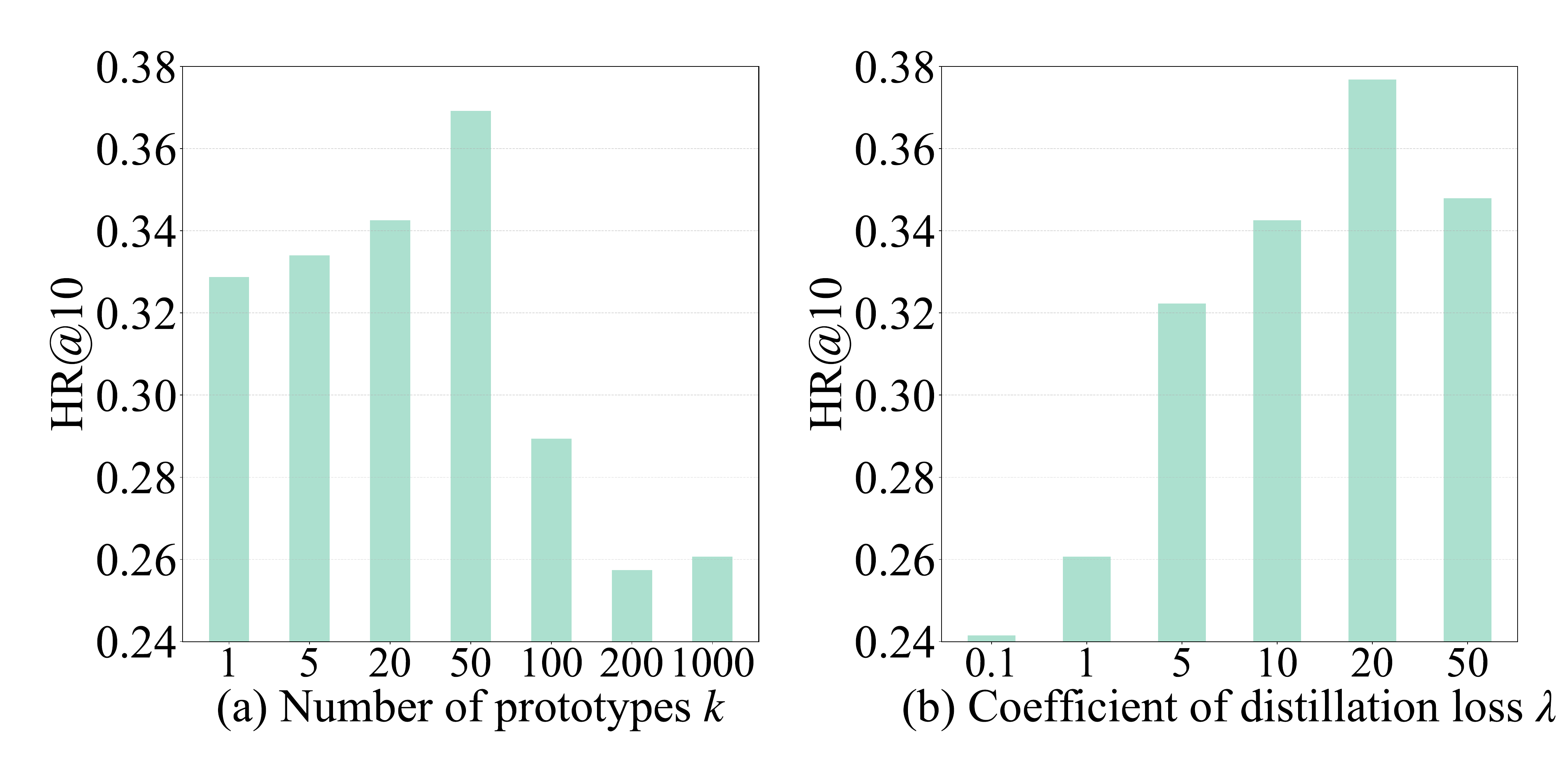}
    \caption{Effect of two hyper-parameters on model performance (HR@10) on the LastFM dataset.}
    \label{hyperparameter_HR_lastfm}
\end{figure}
\begin{figure}[t]
    \centering
    \includegraphics[width=1.0\linewidth]{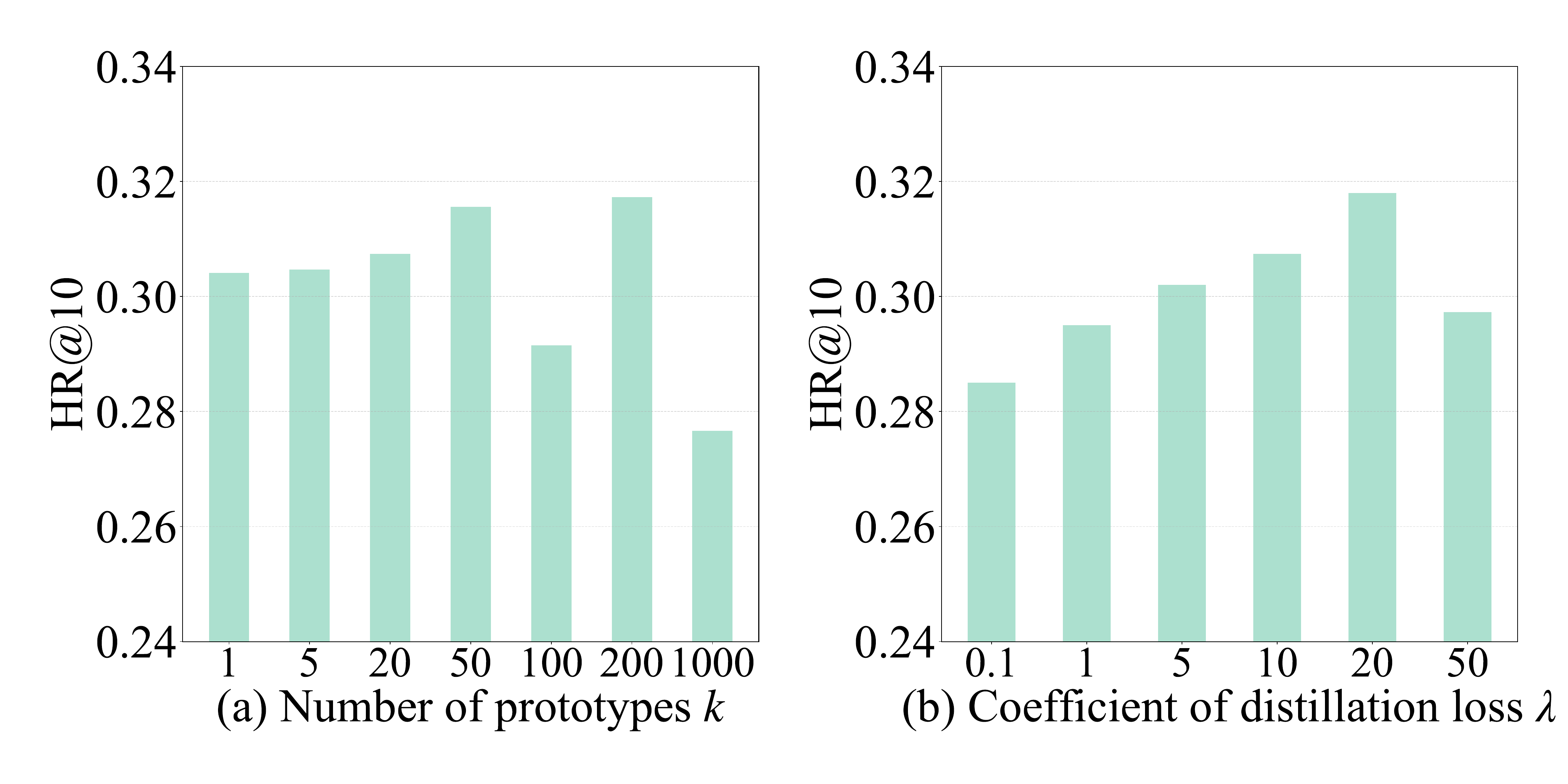}
    \caption{Effect of two hyper-parameters on model performance (HR@10) on the Tmall dataset.}
    \label{hyperparameter_HR_tmall}
\end{figure}
\noindent \textbf{The number of semantically aligned prototypes $k$.} 
We vary $k$ among $\{1, 5, 20, 100, 200, 1000\}$ to evaluate its influence on model performance across \textit{RetailRocket}, \textit{LastFM}, and \textit{Tmall}.
Figures~\ref{hyperparameter_HR_retail}(a) to~\ref{hyperparameter_HR_tmall}(a) show consistent yet dataset-specific trends.
While all datasets exhibit the same non-monotonic pattern as in the XING dataset performance. But there are subtle differences in the turning points.
For instance, the optimal $k$ for RetailRocket and LastFM occurs at smaller values (around 20–50), whereas Tmall benefits from a slightly larger $k$ (around 200).
This suggests that datasets with higher user diversity tend to favor a broader prototype set, reflecting the trade-off between representational diversity and overfitting risk.

\noindent \textbf{The coefficient of distillation loss $\lambda$.} 
We further analyze $\lambda \in \{0.1, 1, 5, 10, 20, 50\}$ to assess its influence on performance stability.
As shown in Figures~\ref{hyperparameter_HR_retail}(b) to~\ref{hyperparameter_HR_tmall}(b), the optimal $\lambda$ values across datasets consistently fall within the range of 10–20, aligning with the findings on the XING dataset in the main text.
However, the sensitivity to $\lambda$ varies across datasets. Tmall shows a more gradual performance drop as $\lambda$ increases, while RetailRocket exhibits a sharper decline when $\lambda > 20$.
These observations imply that datasets with more dynamic user behavior may require stronger temporal regularization to stabilize the learning process.

\section{Results of Practical Viability Analysis}
\subsection{Efficiency Improvement}
To complement the analysis presented in Section~5.5, we further evaluate the proposed efficiency refinement mechanisms on the remaining three datasets, including RetailRocket, LastFM, and Tmall. 

\noindent \textbf{(1) Limiting the size of the knowledge base.}

We apply the same sliding-window strategy described in the main text, where only client prototypes from the most recent $r$ communication rounds are retained. As shown in Table~\ref{tab:KB_restricted}, model performance remains stable even when the knowledge base size is substantially reduced. In LastFM dataset, moderate window sizes yield slightly better performance, suggesting that pruning outdated prototypes can effectively mitigate noise accumulation and enhance prototype relevance.

% 原文Instead of retaining all client prototypes across communication rounds, we adopt a sliding-window strategy that retains only the client prototypes from the most recent communication rounds $r$. Once the number of rounds exceeds $r$, client prototypes from earlier rounds are discarded and replaced with those from newer ones. As shown in Table~\ref{tab:KB_restricted}, reducing the size of the prototype knowledge base maintains stable performance and in some cases even improves it, while significantly lowering computational overhead.

\noindent \textbf{(2) Reducing the frequency of prototype transfer.}

To validate the generality of the findings reported, table~\ref{tab:different intervals} confirm that the same trend holds: increasing the transfer interval within a moderate range preserves model performance while notably reducing communication and computation overhead. This consistency across datasets substantiates the robustness and scalability of the proposed mechanism.

\subsection{Privacy Considerations}

\begin{figure}[!t]
    \centering
    \includegraphics[width=0.8\linewidth]{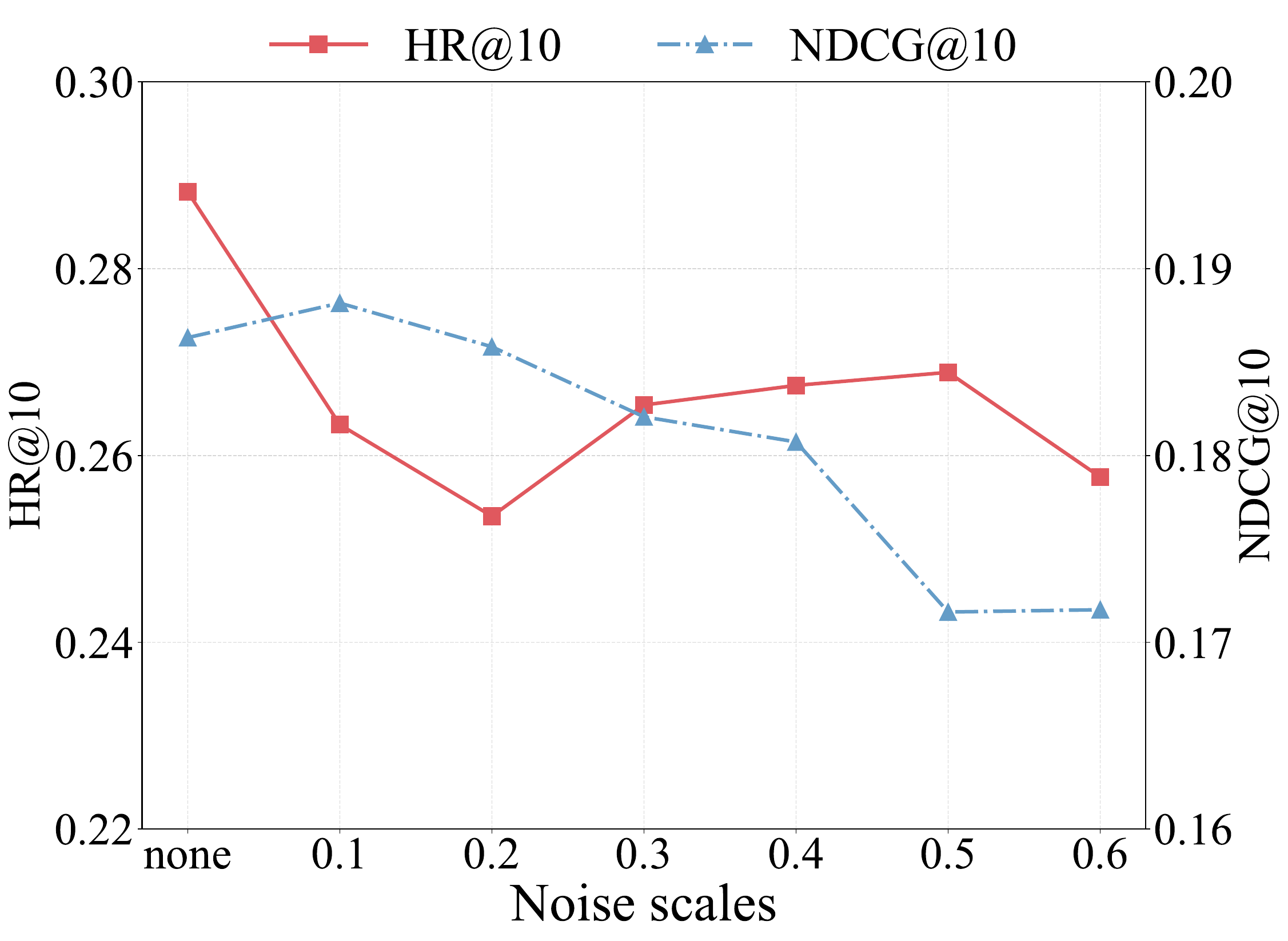}
    \caption{Performance of integrating LDP into our \model on XING dataset.}
    \label{privacy_xing}
\end{figure}

\begin{figure}[!t]
    \centering
    \includegraphics[width=0.8\linewidth]{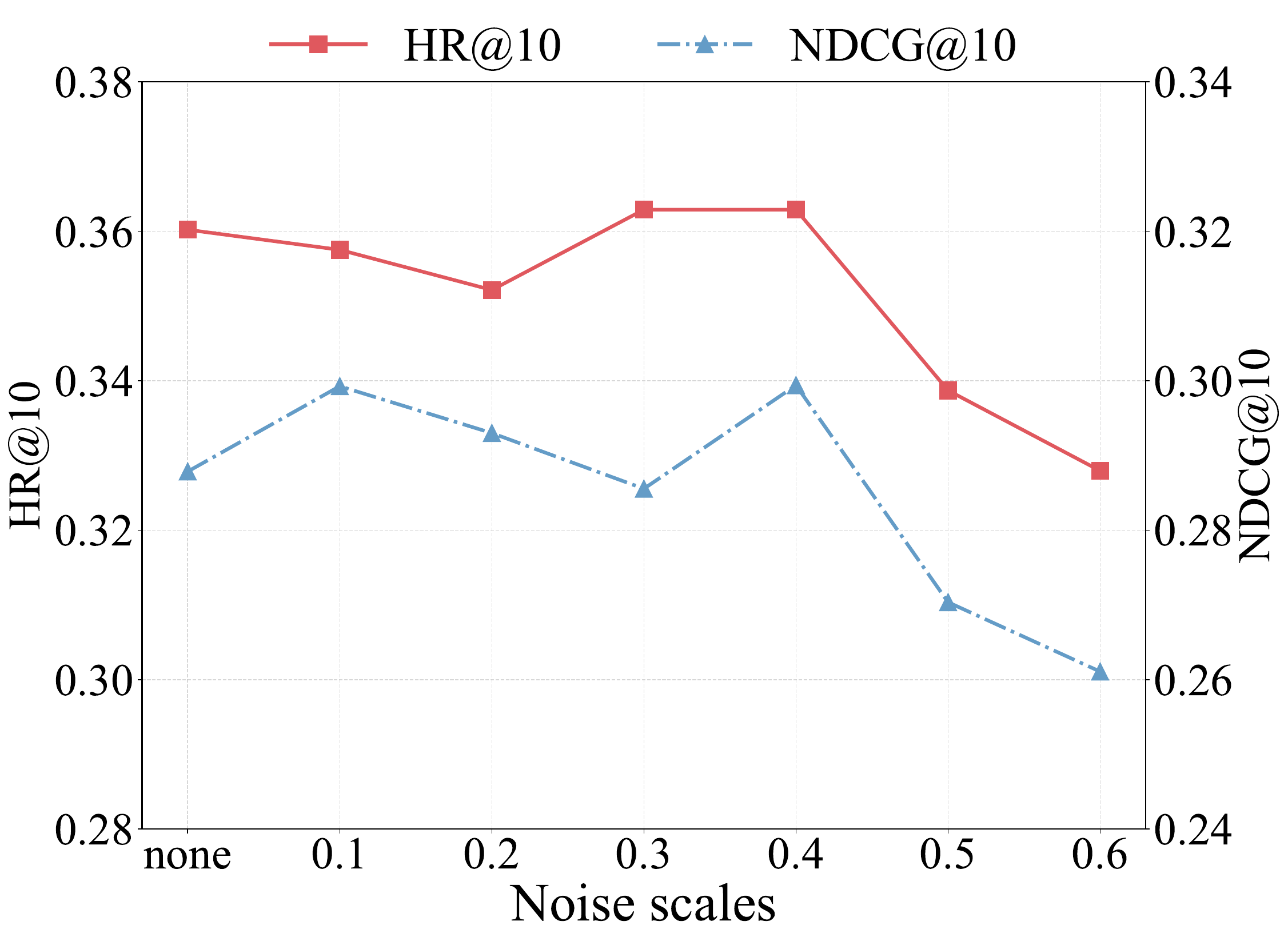}
    \caption{Performance of integrating LDP into our \model on RetailRocket dataset.}
    \label{privacy_retailrocket}
\end{figure}

\begin{figure}[!t]
    \centering
    \includegraphics[width=0.8\linewidth]{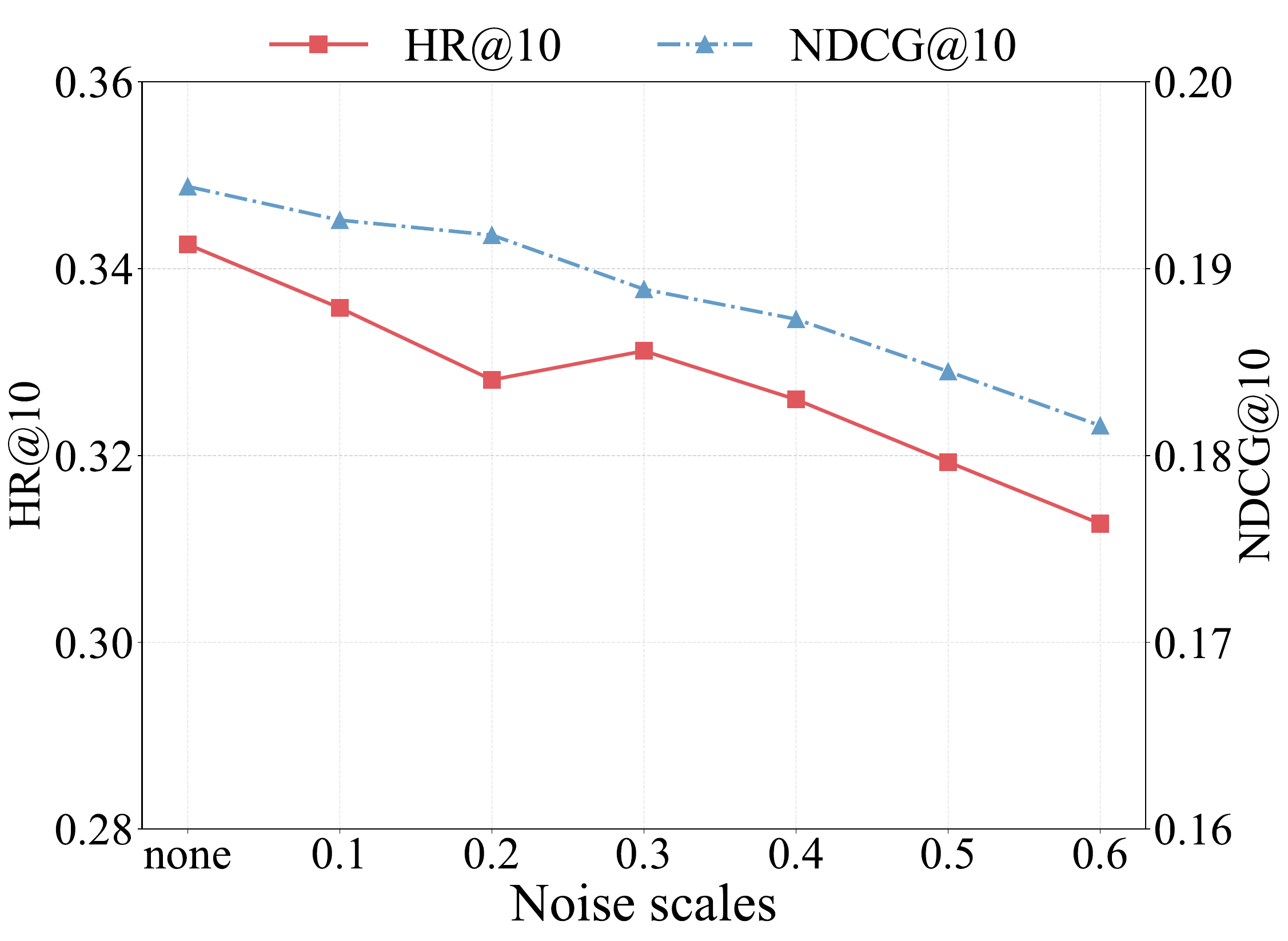}
    \caption{Performance of integrating LDP into our \model on LastFM dataset.}
    \label{privacy_laastfm}
\end{figure}

\begin{figure}[!t]
    \centering
    \includegraphics[width=0.8\linewidth]{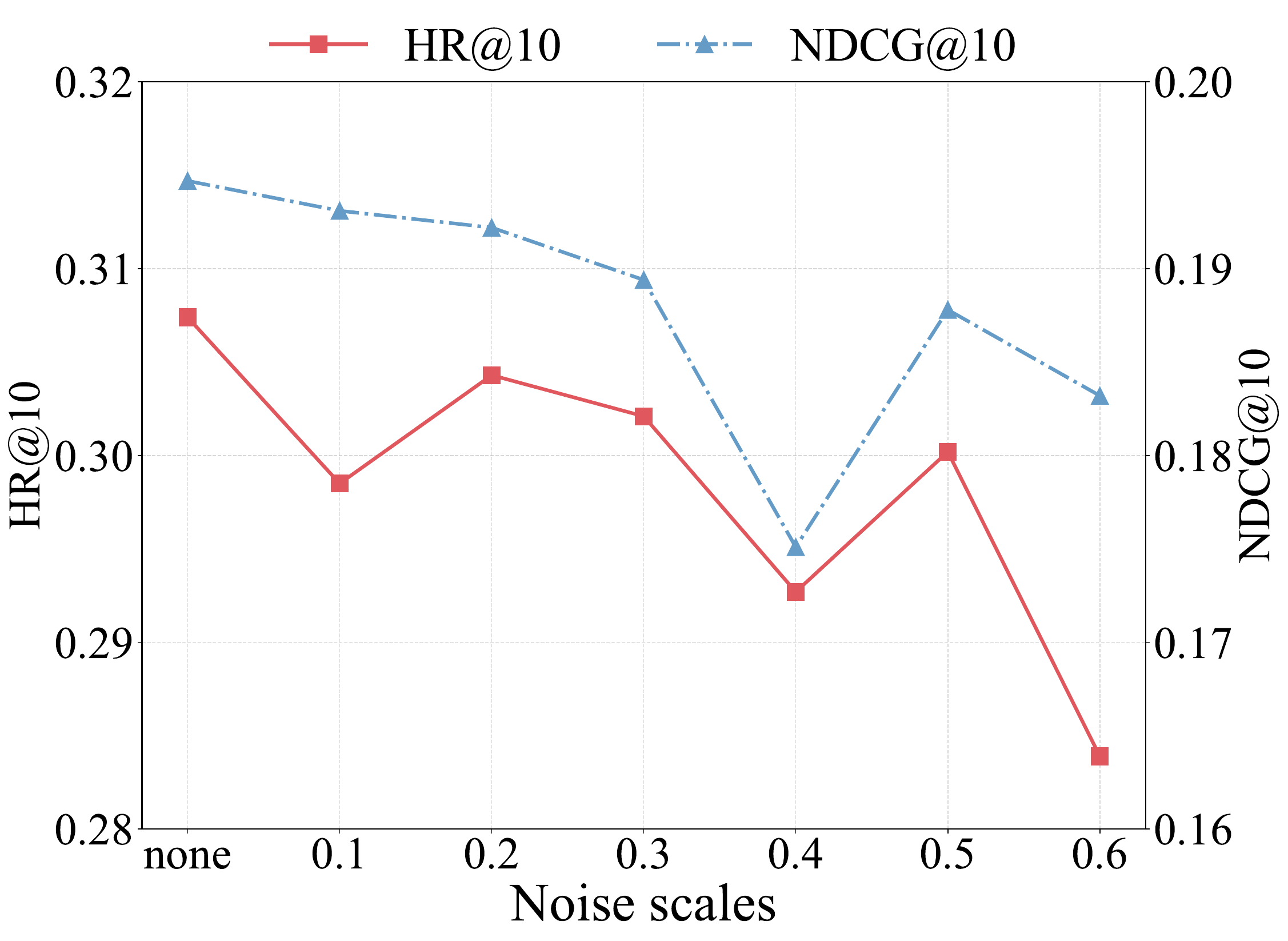}
    \caption{Performance of integrating LDP into our \model on Tmall dataset.}
    \label{privacy_tmall}
\end{figure}

To complement the main results reported in Section~5.6, we further evaluate the privacy robustness of our \model on all datasets. Following the same local differential privacy (LDP) setup as in the main text, we inject Laplacian noise with different standard deviations ($std=\{0.1,\ldots,0.6\}$) into client parameters before aggregation. 
As illustrated in Figure~\ref{privacy_xing} to Figure~\ref{privacy_tmall}, the performance degradation under increasing noise levels remains moderate across all datasets. In particular, when $std \leq 0.3$, \model maintains more than 95\% of its original accuracy while ensuring a reasonable privacy guarantee. These consistent trends across datasets verify that our framework achieves a robust trade-off between privacy preservation and recommendation quality.

% \model is inherently privacy-friendly, as each client transmits a single session-level representation and representation module parameters without any user-identifiable information. This design allows easy integration with privacy-preserving techniques. We further apply local differential privacy (LDP) by injecting Laplacian noise into client parameters, using varying noise scales $std=\{0.1,\ldots,0.6\}$. These correspond to approximately $(\epsilon, \delta)$-LDP with $\delta = 10^{-5}$ and $\epsilon$ ranging from 48.5 (weaker privacy) to 8.1 (stronger privacy), as calculated using the Gaussian mechanism. As shown in Figure~\ref{privacy_retailrocket} to~\ref{privacy_tmall}, \model remains robust under different noise levels, maintaining competitive performance while enhancing privacy protection.

\end{document}